\documentclass[10pt,a4paper]{article}
\usepackage{amsmath}
\usepackage{amsfonts}
\usepackage{amssymb}
\usepackage{graphicx}
\usepackage{textcomp}
\usepackage{url}
\usepackage{lipsum}
\usepackage{setspace}
\usepackage{color}
\usepackage[numbers,sort&compress]{natbib}
\usepackage[usenames,dvipsnames,svgnames,table]{xcolor}
\usepackage[lofdepth,lotdepth,caption=false]{subfig}
\usepackage[pdfstartview=FitH]{hyperref}
\usepackage{hyperref}
\newsavebox{\bigleftbox}
\hypersetup{
 colorlinks,
 citecolor=Blue,
 linkcolor=Blue,
 urlcolor=Blue
}

\makeatletter
 \def\footnoterule{\kern-3\p@
   \noindent\hrulefill \kern 2.8\p@} % the \hrule is .4pt high
\makeatother
\usepackage[left=3.00cm, right=3.00cm, top=2.00cm, bottom=2.00cm]{geometry}

\usepackage{fancyhdr}

\title{\textbf{Thermal Stability and Fracture Patterns of a Recently Synthesized Monolayer Fullerene Network: A Reactive Molecular Dynamics Study}}
\author{
	L. A. Ribeiro Junior$^{1,\dag}$, 
	M. L. Pereira Júnior$^{2,\ddag}$,
	W. F. Giozza$^{2,*}$,\\
	R. M. Tromer$^{3,4,**}$, and
	D. S. Galvão$^{3,4,\S}$
	}
\date{}

\begin{document}
    \maketitle
	\vspace{-0.6cm}
	\begin{center}\small
		$^1$\textit{Institute of Physics, University of Brasília, Brasília, Brazil}\\
		$^2$\textit{Department of Electrical Engineering, Faculty of Technology, University of Bras\'{i}lia, Bras\'{i}lia, Brazil}\\
		$^3$\textit{Center for Computing in Engineering \& Sciences, Unicamp, Campinas, SP, Brazil}\\
		$^4$\textit{Applied Physics Department, `Gleb Wataghin' Institute of Physics, Unicamp, Campinas, SP, Brazil}\\ \vspace{0.2cm} 
		\phantom{.}\hfill
		$^{\dag}$\url{ribeirojr@unb.br}\hfill
		$^{\ddag}$\url{marcelo.lopes@unb.br}\hfill
		$^*$\url{giozza@unb.br}\hfill
		\phantom{.}
		\\
		\phantom{.}\hfill
		$^{**}$\url{rrrtremer@gmail.com}\hfill
		$^{\S}$\url{galvao@ifi.unicamp.br}\hfill
		\phantom{.}
	\end{center}
	
% ---------------------- %
%\blfootnote{}
%\blfootnote{}

\onehalfspace

\noindent\textbf{Abstract:} New monolayer 2D carbon structures, namely qHPC$_{60}$ and qTPC$_{60}$, were recently synthesized by covalently bonding C$_{60}$ polymers. Here, we carried out Reactive (ReaxFF) molecular dynamics simulations to study the thermodynamic stability and fracture patterns of qHPC$_{60}$ and qTPC$_{60}$. Our results showed that these structures present similar thermal stability, with sublimation points of 3898K and 3965K, respectively. qHPC$_{60}$ and qTPC$_{60}$ undergo an abrupt structural transition becoming totally fractured after a critical strain threshold. The crack propagation is linear (non-linear) for qHPC$_{60}$ (qTPC$_{60}$). The estimated elastic modulus for qHPC$_{60}$ and qTPC$_{60}$ are 175.9 GPa and 100.7 GPa, respectively.   

\section{Introduction}

Since the discovery of graphene \cite{novoselov2004electric}, 2D carbon materials have been widely studied to develop new optoelectronic applications \cite{kumar2018recent}. Their controllable synthesis can yield different structures with distinct physical and chemical properties \cite{lu2013two,wang2016electronic}. Due to this reason, the large variety of 2D carbon allotropes proposed so far \cite{enyashin2011graphene,wang2015phagraphene,wang2018popgraphene,zhuo2020me,karaush2014dft,zhang2019art,toh2020synthesis,fan2021biphenylene,PhysRevB.70.085417,Alsayoud2018} have proven to be suitable for flat electronics, although only a few structures has been experimentally realized. 

Recently, new 2D carbon allotropes were synthesised: the monolayer amorphous carbon (MAC) \cite{toh2020synthesis} and the 2D biphenylene network (BPN) \cite{fan2021biphenylene}. MAC is made of randomly distributed five, six, seven, and eight atom rings. BPN consists of a periodic lattice composed of fused rings containing four, six, and eight carbon atoms. Similar to graphene, MAC and BPN present a zero semi-metal bandgap, which is a serious drawback to their usage in some optoelectronic applications \cite{raccichini2015role}.  

Very recently, 2D carbon materials with a semiconducting bandgap of about 1.6 eV --- namely monolayer quasi-hexagonal-phase fullerene (qHPC$_{60}$) and monolayer quasi-tetragonal-phase fullerene qTPC$_{60}$) --- were experimentally realized overcoming the problem of a null bandgap shown by other 2D carbon-based materials \cite{hou2022synthesis}. It was reported that qHPC$_{60}$ and qTPC$_{60}$ show excellent environmental and thermal stabilities. The Raman spectra and optical images remain unchanged after heating, indicating that the polymeric C$_{60}$ frameworks do not decompose up to 600K. qHPC$_{60}$ and qTPC$_{60}$ exhibit high crystallinity and unique topological structure \cite{hou2022synthesis}. In this sense, a detailed description of their structural integrity with a focus on thermodynamic stability and stress resilience can pave the way for its applications. 

In the present work, we investigated the thermodynamic stability and stress resilience of qHPC$_{60}$ and qTPC$_{60}$. It was observed that these materials present a well-defined (linear) elastic region when subjected to uniaxial strain. qHPC$_{60}$ and qTPC$_{60}$ undergo an abrupt structural transition to a totally fractured form after a critical strain threshold. The crack propagation is linear for qHPC$_{60}$ and non-linear for qTPC$_{60}$, respectively. Heating ramp protocol simulations revealed that they are thermally stable up to 3898K and 3965K, respectively. 

\section{Methodology}

\begin{figure}[!htb]
	\centering
	\includegraphics[width=\linewidth]{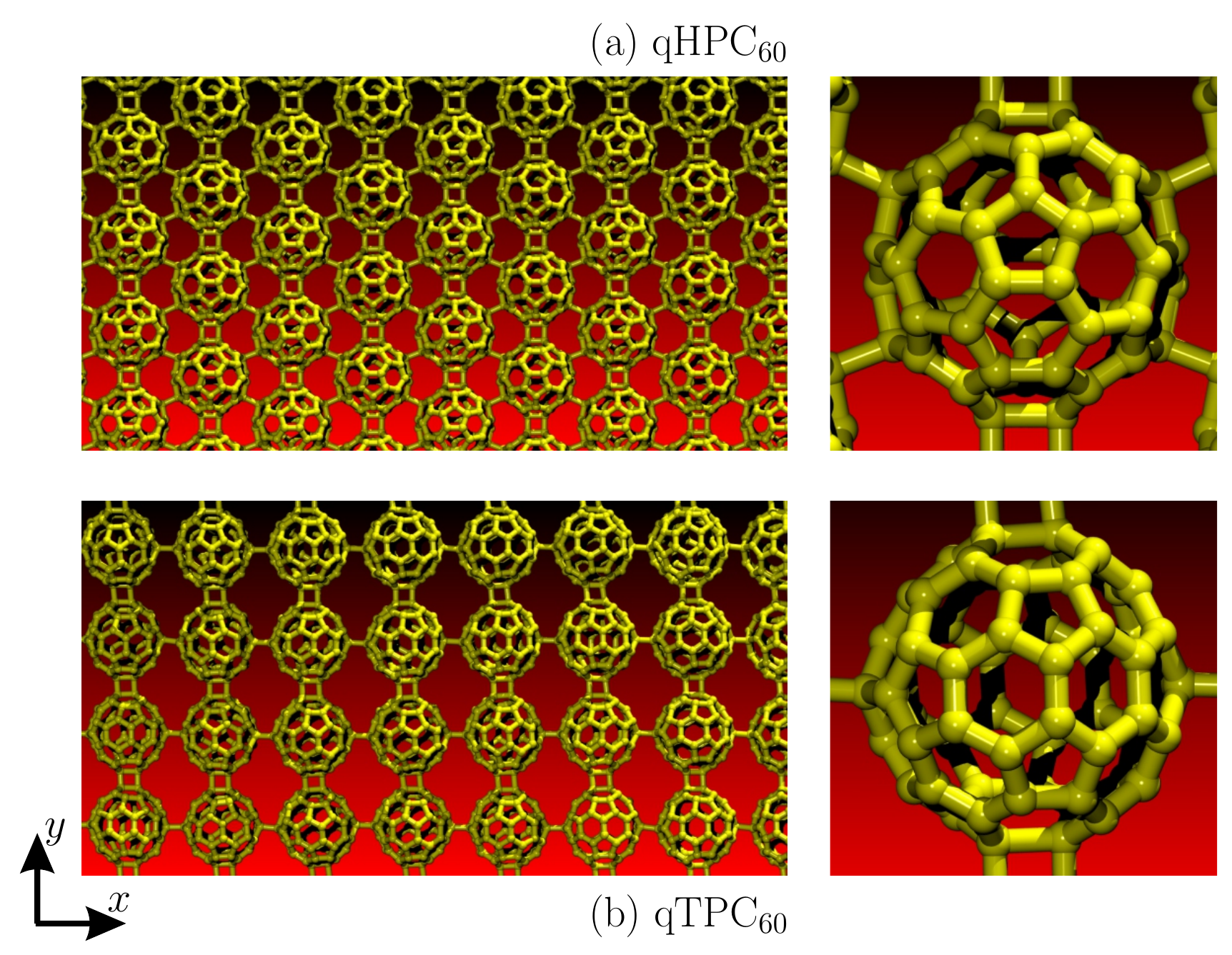}
	\caption{Schematic representation of: (a) qHPC$_{60}$ and (b) qTPC$_{60}$ lattice structures. The right panels illustrate the corresponding unit cells.}
	\label{fig-system}
\end{figure}

We carried out fully-atomistic MD simulations with the reactive force field ReaxFF (by employing the parameter set for C/H/O \cite{senftle2016reaxff,smith2017reaxff}), as implemented in LAMMPS \cite{plimpton_JCP}. A reactive potential is needed (it allows the formation and breaking of chemical bonds during the dynamics) for the fracture dynamics investigation. Figure \ref{fig-system} illustrates the qHPC$_{60}$ and qTPC$_{60}$ structural models (supercells) used in the MD simulations. Both structures have dimensions of $100\times 100$ \r{A}${^2}$ with periodic boundary conditions and are composed of 8640 atoms. The length of the simulation box along the z-direction is 200 \r{A}.

The equations of motion were numerically integrated using the velocity-Verlet algorithm, with a time-step of $0.2$ fs. The uniaxial tensile strain was applied along the periodic x and y directions, for an engineering strain rate of $1.0 \times 10^{-6}$ fs$^{-1}$. Before their heating and stretching, the structures were equilibrated using an NPT ensemble at a constant temperature of 300 K and null pressures using a Nos\'e-Hoover thermostat \cite{hoover1985canonical} during 200 ps. 

To obtain the mechanical properties and fracture patterns, the qHPC$_{60}$ and qTPC$_{60}$ structures were continuously stretched up to their complete structural failure (fracture) by applying a maximum strain of 50\%. The tensile stretching simulations were performed by increasing the x and y cell dimensions. Their thermal stability was investigated by heating up the structures from 300K up to 10000K. The heating process is simulated by linearly increasing the temperature during 1 ns in an NVT ensemble. The MD snapshots and trajectories were obtained using the visualization and analysis software VMD \cite{HUMPHREY199633}.

\section{Results}

\begin{figure}[!htb]
    \centering
    \includegraphics[width=\linewidth]{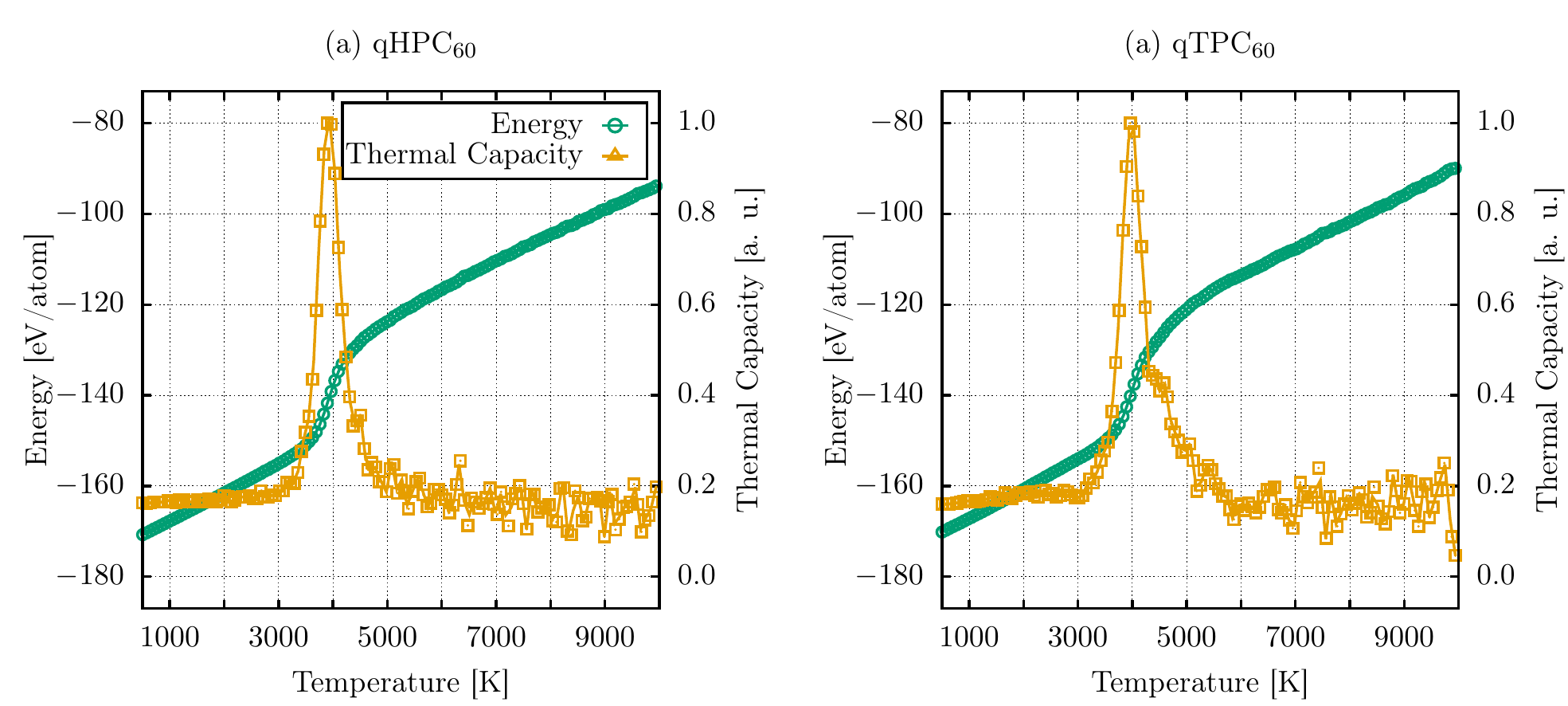}
	\caption{Total energy and heat capacity ($C_V$) as a function of temperature for: (a) qHPC$_{60}$ and (b) qTPC$_{60}$ at 1 ns of the heating ramp simulations.}
    \label{fig-melting-curves}
\end{figure}

\begin{figure}[!htb]
	\centering
	\includegraphics[width=\linewidth]{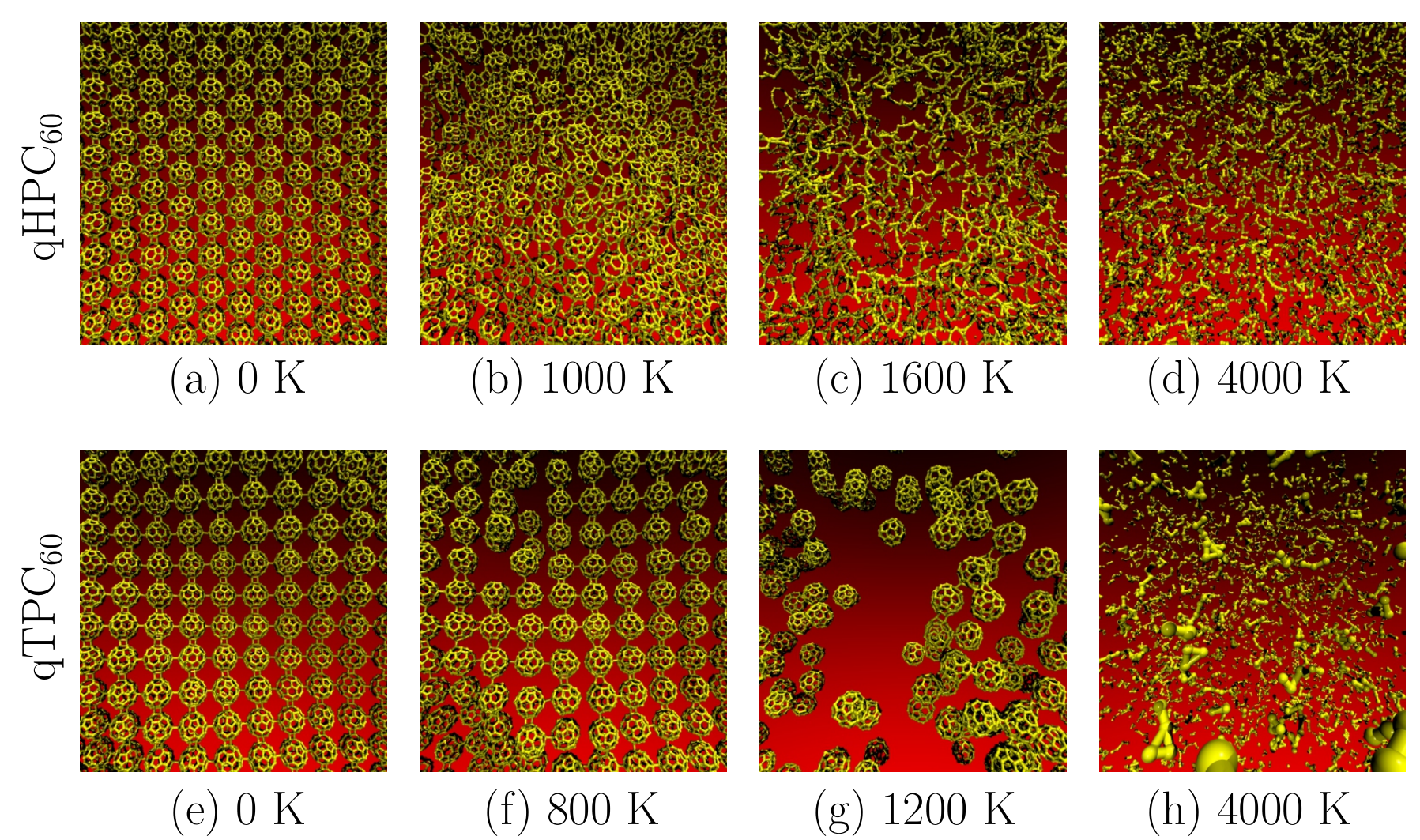}
	\caption{Representative MD snapshots for the heating ramp simulations (melting process) for: qHPC$_{60}$ at (a) 0K, (b) 1000K, (c) 1600K, and (d) 4000K, and qTPC$_{60}$ at (e) 0K, (f) 800K, (g) 1200K, and (h) 4000K.}
	\label{fig-melting}
\end{figure}

We begin our discussion by analyzing the qHPC$_{60}$ and qTPC$_{60}$ thermodynamic stability. The melting processes occur within the heating ramp simulations, as mentioned above. Figure \ref{fig-melting} illustrates the total energy (green) and heat capacity ($C_V$, in yellow) as a function of temperature for the melting process of qHPC$_{60}$ (Figure \ref{fig-melting-curves}(a)) and qTPC$_{60}$ (Figure \ref{fig-melting-curves}(b)). As a general trend, we can see that the total energy increases quasi-linearly with the temperature showing three different slopes: between 300K-3500K, 3500K-4000K, and 4000K-10000K. 

qHPC$_{60}$ and qTPC$_{60}$ maintain their structural integrity in the first stage of the heating process, defined by temperatures up to 3500K. After this critical value, the thermal vibrations impose substantial changes in their morphologies, and the melting process occurs in the second stage of the heating process (for temperatures ranging from 3500K up to 4500K). The melting point is defined by the most pronounced peak in the $C_V$ curves. In this sense, the peak in the heat capacity curve indicates the similar melting points at 3898K (see Figure \ref{fig-melting-curves}(a)) and 3965K (see Figure \ref{fig-melting-curves}(b)) for qHPC$_{60}$ and qTPC$_{60}$, respectively.   

The third stage of the heating process (between 4000K-10000K) in Figures \ref{fig-melting-curves}(a) and \ref{fig-melting-curves}(b) characterizes the continuous heating of the systems. The abrupt change in the slope for the total energy curves defines a phase transition from a solid to a gas-like phase (sublimation) of the carbon atoms. The melting points obtained here for qHPC$_{60}$ and qTPC$_{60}$ are comparable to those for the monolayer graphene (4095K) \cite{los2015melting}, MAC (3626K) \cite{felix2020mechanical}, and BPN (4024K) \cite{pereira2022mechanical}.

We further explored the thermodynamic stability of qHPC$_{60}$ and qTPC$_{60}$ by analyzing the representative MD snapshots for the heating ramp simulations, with temperatures varying from 300K up to 4000K, as shown in Figure \ref{fig-melting}. Figures \ref{fig-melting-curves}(a) and \ref{fig-melting-curves}(e) illustrate the lattice structure for qHPC$_{60}$ and qTPC$_{60}$ at 0K, respectively. In Figures \ref{fig-melting-curves}(b-c) and \ref{fig-melting-curves}(f-g), we can note the melting process of these structures occurs differently. For the qHPC$_{60}$, the C$_{60}$ units are fragmented into several linear atomic chains (LACs) at 1600K, while the qTPC$_{60}$ tends to fragment into separated C$_{60}$ units at 1200K. 

The complete atomization of the structures occurs at the related melting points, as shown in Figures \ref{fig-melting-curves}(d) and \ref{fig-melting-curves}(h) for qHPC$_{60}$ and qTPC$_{60}$, respectively. Importantly, the different mechanisms observed in the melting processes are strictly related to their topology. qHPC$_{60}$ presents eight covalent bonds connecting the next-neighboring C$_{60}$ units, whereas qTPC$_{60}$ presents only six. The higher number of covalent bonds in qHPC$_{60}$ favors the formation of LACs before the lattice atomization at 1000K (see Figure \ref{fig-melting-curves}(b)). Conversely, the qTPC$_{60}$ lower number of covalent bonds is responsible for its fragmentation into individual C$_{60}$ molecules at 800K (see Figure \ref{fig-melting-curves}(f)).    

\begin{figure}[!htb]
\centering
\includegraphics[width=\linewidth]{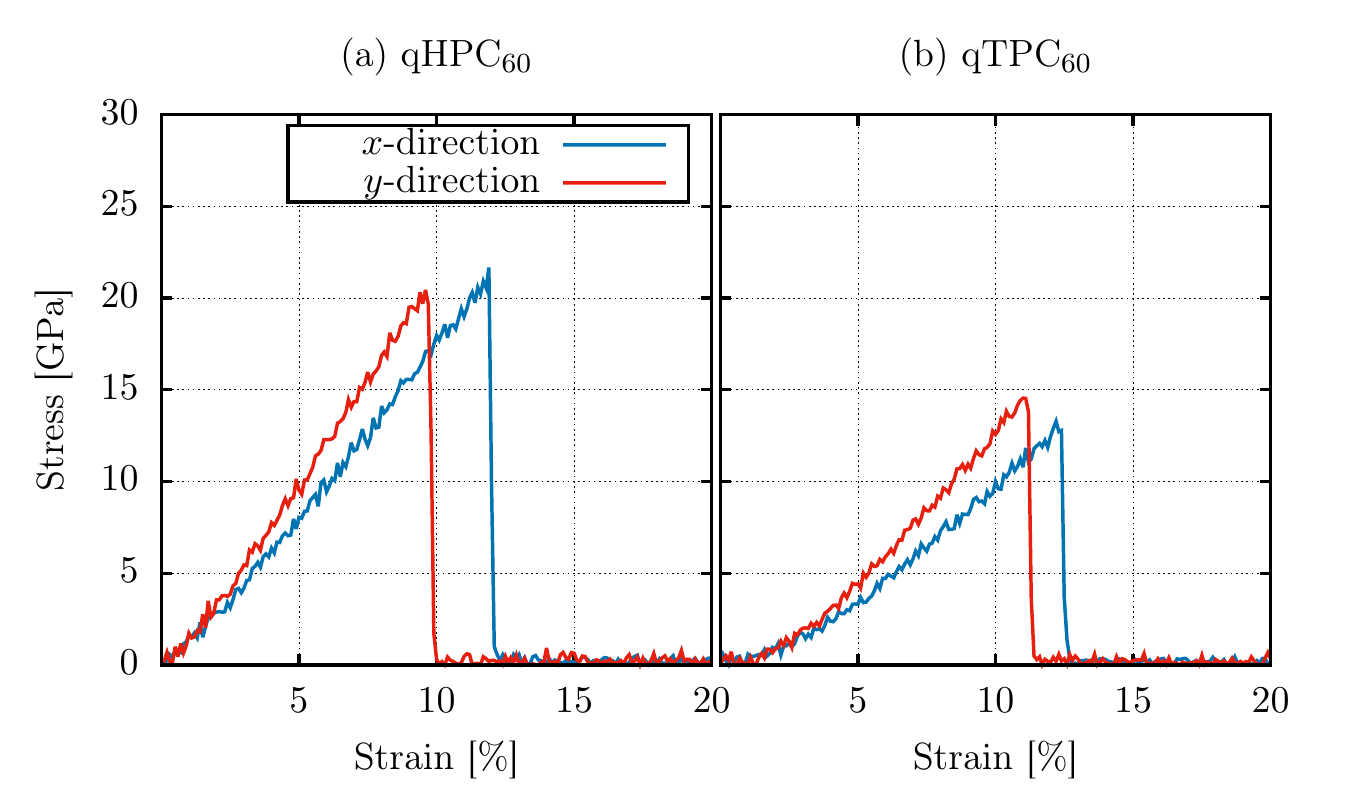}
\caption{Stress-strain curves for (a) qHPC$_{60}$ and (b) qTPC$_{60}$ as a function of the uniaxial applied strain in the (blue) x-direction and (red) y-direction.}
\label{fig-strain-curves}
\end{figure}

The stress response as a function of the uniaxial applied strain is illustrated in Figure \ref{fig-strain-curves}. Figures \ref{fig-strain-curves}(a) and \ref{fig-strain-curves}(b) show the stress-strain curves for the uniaxial tensile loading along the x-direction (blue) and y-direction (red) for qHPC$_{60}$ and qTPC$_{60}$, respectively. The first noticeable feature is that these structures have an anisotropic mechanical behavior, which is expected because of their structural anisotropy. qHPC$_{60}$ and qTPC$_{60}$ are more resilient to tension along the x-direction than for the y-one. The presence of a ring composed of four carbon atoms linking the C$_{60}$ units makes these structures less resilient to tension along the y-direction. 

In Figure \ref{fig-strain-curves}, we can note that these materials present a well-defined (linear) elastic region when subjected to uniaxial strain. qHPC$_{60}$ and qTPC$_{60}$ undergo an abrupt transition from integrity to a fractured form after a critical strain of 9.6\% (11.9\%) along the x-direction (y-direction) and 12.2\% (11.0\%) along the x-direction (y-direction), respectively. The ultimate stress values for qHPC$_{60}$ and qTPC$_{60}$ are: 20.4 GPa (21.7 GPa) for the x-direction (y-direction) and 13.3 GPa (14.6 GPa) for the x-direction (y-direction), respectively. The ultimate stress is defined as the corresponding tensile stress for a critical strain. In our calculations, 1\% of strain was used to estimate the following Young's modulus values for qHPC$_{60}$ and qTPC$_{60}$: 175.9 GPa (218.5 GPa) for the x-direction (y-direction) and 100.7 GPa (133.5 GPa) for the x-direction (y-direction), respectively. As mentioned above, qHPC$_{60}$ tends to be more resilient than qTPC$_{60}$ due to the higher number of covalent bonds connecting the next-neighboring C$_{60}$ units. These values are much lower than other 2D carbon-based structures and can be explained by the high structural stability of the C$_{60}$ units.

\begin{figure}[!htb]
\centering
\includegraphics[width=\linewidth]{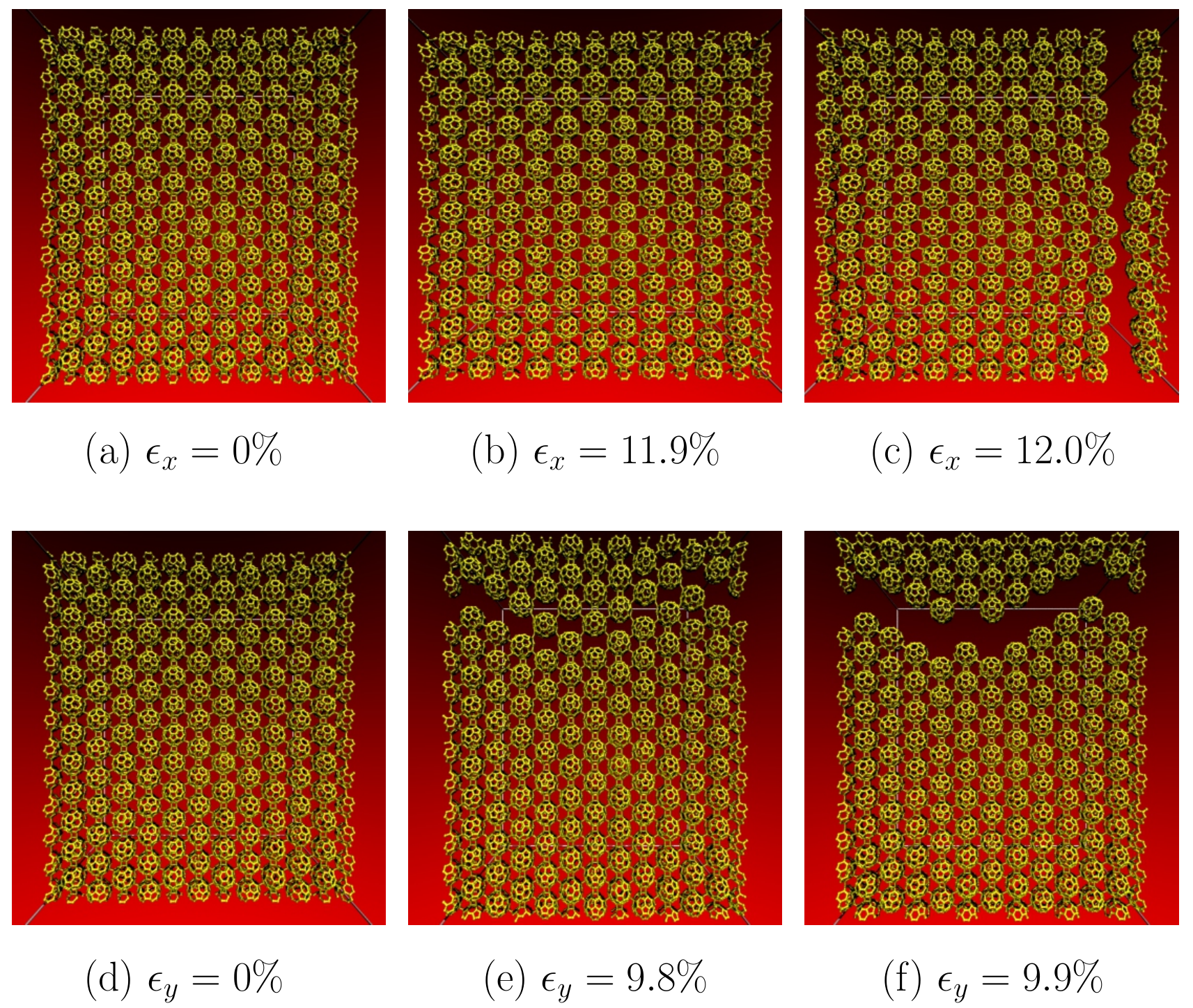}
\caption{Representative MD snapshots for the strain applied along the (a-c) x-direction ($\varepsilon_x$) and (d-f) y-direction ($\varepsilon_y$) for the qHPC$_{60}$ monolayer.}
\label{fig-qhpc60-strain}
\end{figure}

Finally, we discuss the fracture patterns of the qHPC$_{60}$ and qTPC$_{60}$ monolayer when subjected to uniaxial tensile loading. Figures \ref{fig-qhpc60-strain}(a-c) and \ref{fig-qhpc60-strain}(d-f) show representative MD snapshots for the strain applied along the x-direction ($\varepsilon_x$) and y-direction ($\varepsilon_y$) for the qHPC$_{60}$ monolayer. Figure \ref{fig-qhpc60-strain}(a) and \ref{fig-qhpc60-strain}(d) illustrate the monolayer configurations at $\varepsilon_x=0$ and $\varepsilon_y=0$, respectively. As can be inferred from Figure \ref{fig-qhpc60-strain}(b), qHPC$_{60}$ preserves its topology as the strain increase up to  $\varepsilon_x=11.9\%$. Immediately after this critical strain value, the lattice breaks abruptly with fast and linear crack propagation along the opposite direction of the stretch (see Figure \ref{fig-qhpc60-strain}(c)). 

For the qHPC$_{60}$ stretching along the y-direction, a non-linear fast crack propagation is observed along the opposite direction of the stretch at 9.8\% of strain, as shown in Figure \ref{fig-qhpc60-strain}(e). The total fracture of qHPC$_{60}$ is achieved at 9.9 \% of strain (see Figure \ref{fig-qhpc60-strain}(f)). The anisotropic trend for the fracture patterns of qHPC$_{60}$ is due to the different bond arrangements that connect the next-neighboring C$_{60}$ units along the directions of this 2D crystal. For the x-direction, four diagonal bonds connect a central unit to four neighboring units. On the other hand, along the y-direction, there are four parallel bonds linking only two neighboring units with a central one. 

\begin{figure}[!htb]
\centering
\includegraphics[width=\linewidth]{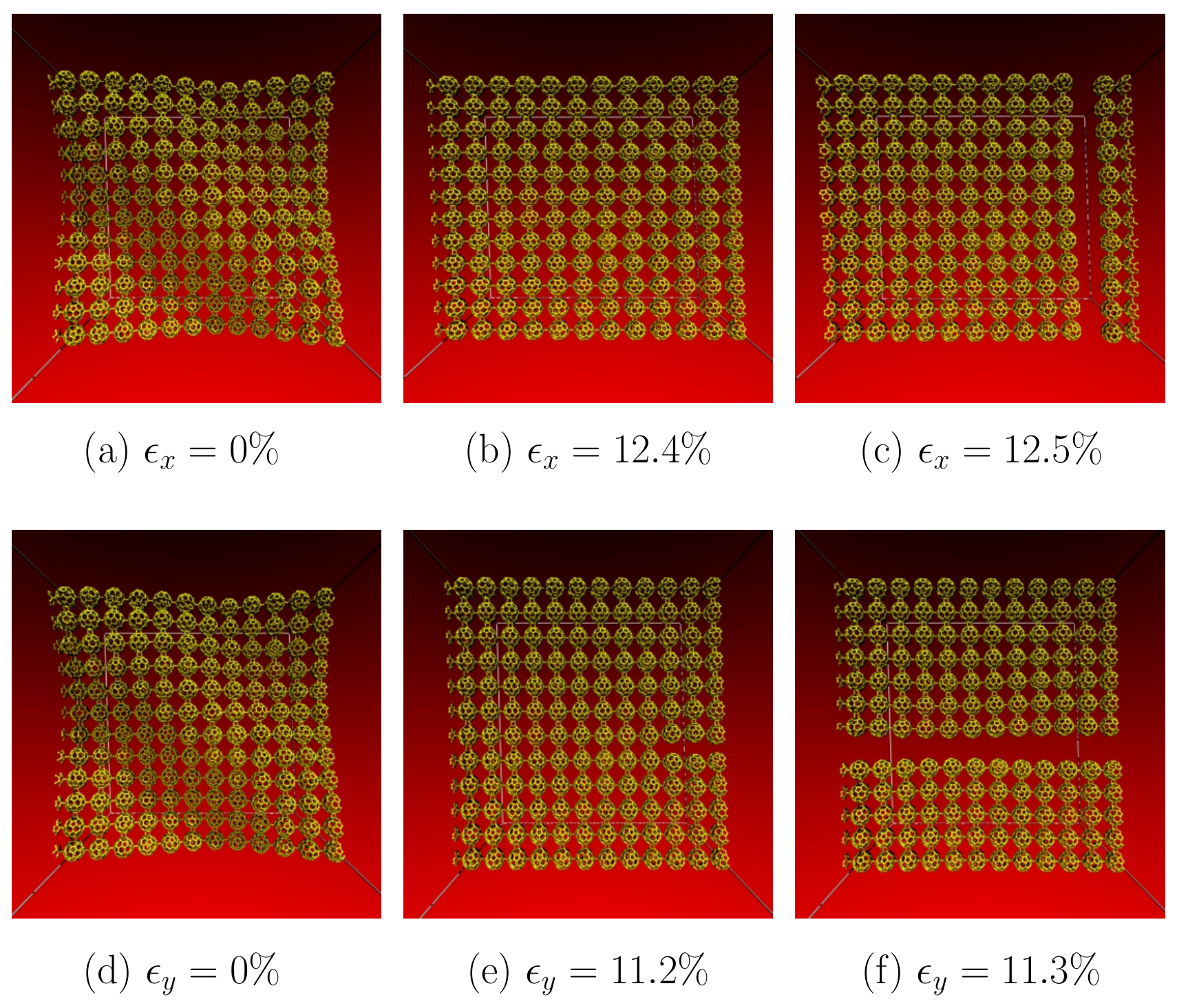}
\caption{Representative MD snapshots for the strain applied along the (a-c) x-direction ($\varepsilon_x$) and (d-f) y-direction ($\varepsilon_y$) for the qTPC$_{60}$ monolayer.}
\label{fig-qtpc60-strain}
\end{figure}

Figures \ref{fig-qtpc60-strain}(a-c) and \ref{fig-qtpc60-strain}(d-f) show representative MD snapshots for the strain applied along the x-direction ($\varepsilon_x$) and y-direction ($\varepsilon_y$) for the qTPC$_{60}$ monolayer. Figure \ref{fig-qtpc60-strain}(a) and \ref{fig-qtpc60-strain}(d) depict the monolayer configurations at $\varepsilon_x=0$ and $\varepsilon_y=0$, respectively. We can note that the equilibrated qTPC$_{60}$ monolayer is not planar. It preserves its topology as the strain increase up to $\varepsilon_x=12.4\%$ (Figure \ref{fig-qhpc60-strain}(b)) and to $\varepsilon_x=11.2\%$ (see Figure \ref{fig-qhpc60-strain}(e)). Immediately after these critical strain values, the lattice breaks linearly with a fast crack propagation along the opposite direction of the stretch for $\varepsilon_x=12.5\%$ (Figure \ref{fig-qhpc60-strain}(c)) and $\varepsilon_y=11.3\%$ (Figure \ref{fig-qhpc60-strain}(f)). The same abrupt fracture mechanism occurs when the qTPC$_{60}$ is stretched along the x and y directions. Since the central C$_{60}$ unity in qTPC$_{60}$ is connected only with parallel covalent bonds to its next-neighboring units, the fracture pattern is the same regardless of the crystal direction.

\section{Conclusions}

In summary, We have carried out fully-atomistic reactive (ReaxFF) MD simulations to investigate the thermodynamic stability and stress resilience of qHPC$_{60}$ and qTPC$_{60}$, which were synthesized very recently \cite{hou2022synthesis}. These structures are 2D carbon-based materials composed of closely packed quasi-hexagonal and quasi-tetragonal crystalline phases of C$_{60}$ molecules. 

Their melting processes were investigated by employing heating ramp simulations, with temperatures varying from 300K to 10000K during one ns. qHPC$_{60}$ and qTPC$_{60}$ maintain their structural integrity up to 3500K. After this critical value, the thermal vibrations impose substantial changes in their morphologies, and the melting process occurs. We obtained similar melting points of 3898K and 3965K for qHPC$_{60}$ and qTPC$_{60}$, respectively. The melting process of these structures occurs differently. For the qHPC$_{60}$, the C$_{60}$ units are fragmented into several linear atomic chains (LACs) at 1600K. The qTPC$_{60}$ structure, in turn, tends to fragment into separated C$_{60}$ units at 1200K.  

For the stress response as a function of the uniaxial applied strain, the first noticeable feature is that qHPC$_{60}$ and qTPC$_{60}$ have a anisotropic mechanical behavior. They are more resilient to tension along the x-direction than along the y-one. The presence of a ring composed of four carbon atoms linking the C$_{60}$ units makes them less resilient to tension along the y-direction. Moreover, these materials present a well-defined (linear) elastic region when subjected to uniaxial tensile loading.  

qHPC$_{60}$ and qTPC$_{60}$ undergo an abrupt structural transition to a fractured form after a critical strain. qHPC$_{60}$ monolayer breaks abruptly with fast and linear crack propagation when stretched along the x-direction. Conversely, non-linear rapid crack propagation was observed by stretching qHPC$_{60}$ along the y-direction. A similar abrupt fracture mechanism occurs when the qTPC$_{60}$ is stretched along the x and y directions. Since a central C$_{60}$ unity in qTPC$_{60}$ is connected only with parallel covalent bonds to its next-neighboring units, the fracture pattern is the same regardless of the crystal direction.   

The following Young's modulus values for qHPC$_{60}$ and qTPC$_{60}$ were estimated: 175.9 GPa (218.5 GPa) for x-direction (y-direction) and 100.7 GPa (133.5 GPa) for x-direction (y-direction), respectively. qHPC$_{60}$ tends to be more resilient than qTPC$_{60}$ due to the higher number of covalent bonds connecting the next-neighboring C$_{60}$ units.

\section*{Acknowledgement}

This work was financed by the Coordenação de Aperfeiçoamento de Pessoal de Nível Superior (CAPES) - Finance Code 001, Conselho Nacional de Desenvolvimento Cientifico e Tecnológico (CNPq), FAP-DF, and FAPESP. We thank the Center for Computing in Engineering and Sciences at Unicamp for financial support through the FAPESP/CEPID Grants \#2013/08293-7 and \#2018/11352-7. L.A.R.J acknowledges the financial support from a FAP-DF grants $00193-00000857/2021-14$ and $00193-00000811/2021-97$, and CNPq grant $302922/2021-0$. W.F.G acknowledges the financial support from a FAP-DF grant $00193-00000853/2021-28$. L.A.R.J. gratefully acknowledges the support from ABIN grant 08/2019 and Fundaç\~ao de Apoio \`a Pesquisa (FUNAPE), Edital 02/2022 - Formul\'ario de Inscriç\~ao N.4. L.A.R.J. acknowledges N\'ucleo de Computaç\~ao de Alto Desempenho (NACAD) and for providing the computational facilities. This work used resources of the Centro Nacional de Processamento de Alto Desempenho em São Paulo (CENAPAD-SP). 

\bibliographystyle{unsrt}
\bibliography{bibliography.bib}

\begin{thebibliography}{10}

\bibitem{novoselov2004electric}
Kostya~S Novoselov, Andre~K Geim, Sergei~V Morozov, De-eng Jiang, Yanshui
  Zhang, Sergey~V Dubonos, Irina~V Grigorieva, and Alexandr~A Firsov.
\newblock Electric field effect in atomically thin carbon films.
\newblock {\em science}, 306(5696):666--669, 2004.

\bibitem{kumar2018recent}
Rajesh Kumar, Ednan Joanni, Rajesh~K Singh, Dinesh~P Singh, and Stanislav~A
  Moshkalev.
\newblock Recent advances in the synthesis and modification of carbon-based 2d
  materials for application in energy conversion and storage.
\newblock {\em Progress in Energy and Combustion Science}, 67:115--157, 2018.

\bibitem{lu2013two}
Haigang Lu and Si-Dian Li.
\newblock Two-dimensional carbon allotropes from graphene to graphyne.
\newblock {\em Journal of Materials Chemistry C}, 1(23):3677--3680, 2013.

\bibitem{wang2016electronic}
Zhanyu Wang, F~Dong, Bo~Shen, RJ~Zhang, YX~Zheng, LY~Chen, SY~Wang, CZ~Wang,
  KM~Ho, Yuan-Jia Fan, et~al.
\newblock Electronic and optical properties of novel carbon allotropes.
\newblock {\em Carbon}, 101:77--85, 2016.

\bibitem{enyashin2011graphene}
Andrey~N Enyashin and Alexander~L Ivanovskii.
\newblock Graphene allotropes.
\newblock {\em physica status solidi (b)}, 248(8):1879--1883, 2011.

\bibitem{wang2015phagraphene}
Zhenhai Wang, Xiang-Feng Zhou, Xiaoming Zhang, Qiang Zhu, Huafeng Dong, Mingwen
  Zhao, and Artem~R Oganov.
\newblock Phagraphene: a low-energy graphene allotrope composed of 5--6--7
  carbon rings with distorted dirac cones.
\newblock {\em Nano letters}, 15(9):6182--6186, 2015.

\bibitem{wang2018popgraphene}
Shuaiwei Wang, Baocheng Yang, Houyang Chen, and Eli Ruckenstein.
\newblock Popgraphene: a new 2d planar carbon allotrope composed of 5--8--5
  carbon rings for high-performance lithium-ion battery anodes from bottom-up
  programming.
\newblock {\em Journal of Materials Chemistry A}, 6(16):6815--6821, 2018.

\bibitem{zhuo2020me}
Zhiwen Zhuo, Xiaojun Wu, and Jinlong Yang.
\newblock Me-graphene: a graphene allotrope with near zero poisson's ratio,
  sizeable band gap, and high carrier mobility.
\newblock {\em Nanoscale}, 12(37):19359--19366, 2020.

\bibitem{karaush2014dft}
Nataliya~N Karaush, Gleb~V Baryshnikov, and Boris~F Minaev.
\newblock Dft characterization of a new possible graphene allotrope.
\newblock {\em Chemical Physics Letters}, 612:229--233, 2014.

\bibitem{zhang2019art}
Run-Sen Zhang and Jin-Wu Jiang.
\newblock The art of designing carbon allotropes.
\newblock {\em Frontiers of Physics}, 14(1):1--17, 2019.

\bibitem{toh2020synthesis}
Chee-Tat Toh, Hongji Zhang, Junhao Lin, Alexander~S Mayorov, Yun-Peng Wang,
  Carlo~M Orofeo, Darim~Badur Ferry, Henrik Andersen, Nurbek Kakenov, Zenglong
  Guo, et~al.
\newblock Synthesis and properties of free-standing monolayer amorphous carbon.
\newblock {\em Nature}, 577(7789):199--203, 2020.

\bibitem{fan2021biphenylene}
Qitang Fan, Linghao Yan, Matthias~W Tripp, Ond{\v{r}}ej Krej{\v{c}}{\'\i},
  Stavrina Dimosthenous, Stefan~R Kachel, Mengyi Chen, Adam~S Foster, Ulrich
  Koert, Peter Liljeroth, et~al.
\newblock Biphenylene network: A nonbenzenoid carbon allotrope.
\newblock {\em Science}, 372(6544):852--856, 2021.

\bibitem{PhysRevB.70.085417}
Savas Berber, Eiji Osawa, and David Tom\'anek.
\newblock Rigid crystalline phases of polymerized fullerenes.
\newblock {\em Phys. Rev. B}, 70:085417, Aug 2004.

\bibitem{Alsayoud2018}
Abduljabar~Qassem Alsayoud, Venkateswara~Rao Manga, Krishna Muralidharan,
  Joshua Vita, Stefan Bringuier, Keith Runge, and Pierre Deymier.
\newblock Atomistic insights into the effect of polymerization on the
  thermophysical properties of 2-d c60 molecular solids.
\newblock {\em Carbon}, 133:267--274, Jul 2018.

\bibitem{raccichini2015role}
Rinaldo Raccichini, Alberto Varzi, Stefano Passerini, and Bruno Scrosati.
\newblock The role of graphene for electrochemical energy storage.
\newblock {\em Nature materials}, 14(3):271--279, 2015.

\bibitem{hou2022synthesis}
Lingxiang Hou, Xueping Cui, Bo~Guan, Shaozhi Wang, Ruian Li, Yunqi Liu, Daoben
  Zhu, and Jian Zheng.
\newblock Synthesis of a monolayer fullerene network.
\newblock {\em Nature}, 606(7914):507--510, 2022.

\bibitem{senftle2016reaxff}
Thomas~P Senftle, Sungwook Hong, Md~Mahbubul Islam, Sudhir~B Kylasa, Yuanxia
  Zheng, Yun~Kyung Shin, Chad Junkermeier, Roman Engel-Herbert, Michael~J
  Janik, Hasan~Metin Aktulga, et~al.
\newblock The reaxff reactive force-field: development, applications and future
  directions.
\newblock {\em npj Computational Materials}, 2(1):1--14, 2016.

\bibitem{smith2017reaxff}
Roger Smith, Kenny Jolley, Chris Latham, Malcolm Heggie, Adri van Duin, Diana
  van Duin, and Houzheng Wu.
\newblock A reaxff carbon potential for radiation damage studies.
\newblock {\em Nuclear Instruments and Methods in Physics Research Section B:
  Beam Interactions with Materials and Atoms}, 393:49--53, 2017.

\bibitem{plimpton_JCP}
Steve Plimpton.
\newblock Fast parallel algorithms for short-range molecular dynamics.
\newblock {\em J. Comput. Phys.}, 117:1--19, 1995.

\bibitem{hoover1985canonical}
William~G Hoover.
\newblock Canonical dynamics: equilibrium phase-space distributions.
\newblock {\em Physical review A}, 31(3):1695, 1985.

\bibitem{HUMPHREY199633}
William Humphrey, Andrew Dalke, and Klaus Schulten.
\newblock Vmd: Visual molecular dynamics.
\newblock {\em Journal of Molecular Graphics}, 14(1):33 -- 38, 1996.

\bibitem{los2015melting}
JH~Los, KV~Zakharchenko, MI~Katsnelson, and Annalisa Fasolino.
\newblock Melting temperature of graphene.
\newblock {\em Physical Review B}, 91(4):045415, 2015.

\bibitem{felix2020mechanical}
Levi~C Felix, Raphael~M Tromer, Pedro~AS Autreto, Luiz~A Ribeiro~Junior, and
  Douglas~S Galvao.
\newblock On the mechanical properties and thermal stability of a recently
  synthesized monolayer amorphous carbon.
\newblock {\em The Journal of Physical Chemistry C}, 124(27):14855--14860,
  2020.

\bibitem{pereira2022mechanical}
ML~Pereira, WF~da~Cunha, RT~de~Sousa, GD~Amvame Nze, DS~Galv{\~a}o, and
  LA~Ribeiro.
\newblock On the mechanical properties and fracture patterns of the
  nonbenzenoid carbon allotrope (biphenylene network): a reactive molecular
  dynamics study.
\newblock {\em Nanoscale}, 14(8):3200--3211, 2022.

\end{thebibliography}
	
\end{document}